\documentclass[]{AO4ELT}  

\usepackage{microtype}
\usepackage{biblatex}
\usepackage{amsmath,amsfonts,amssymb}
\usepackage{graphicx}
\usepackage{pst-all} 
\usepackage[colorlinks=true, allcolors=blue]{hyperref}
\makeatletter         
\def\@maketitle{
\includegraphics[width = 170mm]{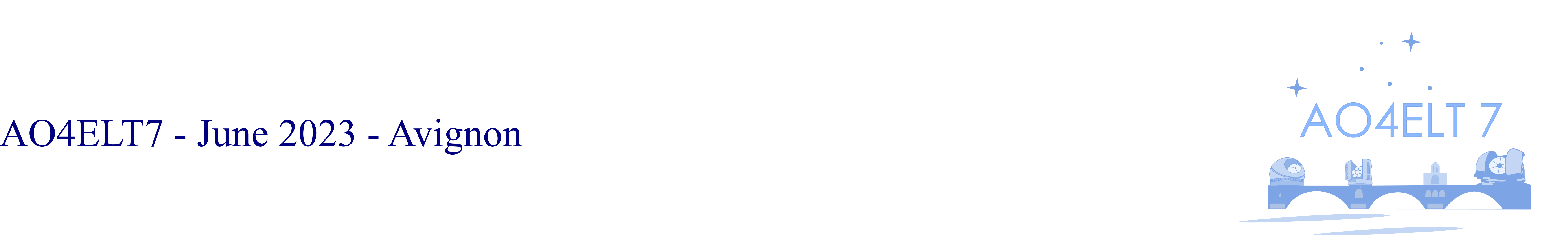}\\[8ex]
\begin{center}
{\Huge \bfseries \sffamily \@title }\\[4ex] 
{\Large  \@author}\\[4ex] 
\@date
\end{center}}

\title{A low-cost, high-speed, very high-order Shack-Hartmann sensor for testing TMT deformable mirrors}

\author[a,b]{Mojtaba Taheri}
\author[c]{David Andersen }
\author[d]{Jean-Pierre Veran}
\author[d]{Olivier Lardière}
\affil[a]{Laboratoire d'Astrophysique de Marseille, Address, 38 Rue Frédéric Joliot Curie, 13013 Marseille, France}
\affil[b]{W.M. Keck Observatory, 65-1120 Mamalahoa Hwy, Waimea, HI, United States}
\affil[c]{TMT International Observatory, 100 W Walnut St, Pasadena, CA , United States}
\affil[d]{NRC Herzberg Astronomy and Astrophysics, 5071 W Saanich Rd, Victoria, BC, Canada}

\begin{document} 
\maketitle
\begin{abstract}
The Thirty Meter Telescope will use a sophisticated adaptive optics system called NFIRAOS. This system utilizes two deformable mirrors conjugate to 0 km and 11.2 km to apply a Multi-Conjugate Adaptive Optics (MCAO) correction over a 2 arcminute field of view. DM0 and DM11 have 63 and 75 actuators across their respective diameters. To study the behavior of these mirrors, we have developed a low-cost, very high-order Shack-Hartmann Wavefront Sensor (WFS). We will use our WFS to calibrate the flatness of the DMs and measure the influence functions of the actuators. NFIRAOS is cooled to reduce the thermal emissivity of optical surfaces visible to the science detectors, so we will also measure the behaviour of the DMs in both warm and cold environments. As the cold chamber is prone to vibrations, a WFS is preferred to a phase-shifting interferometer. Our design was driven by the need to be able to evaluate the DM surface between the actuators, which led to the requirement of at least 248 sub apertures across the diameter. The largest commercially available Shack-Hartmann WFS has only 128 sub-apertures across the diameter, which is not enough to properly sample these DMs. Furthermore, the designed sensor is able to record the wavefront at 50 FPS (50 times per second) at full resolution. To fabricate this WFS, we used a commercial off-the-shelf CMOS detector, camera lens, and lens let array, which kept the total cost less than 20K USD. Here we present the design and performance characteristics of this device.
\end{abstract}


\section{INTRODUCTION} \label{chp:ssh}
Shack-Hartmann Wavefront Sensors (SHWFSs) are the oldest and most frequently used WFSs at astronomical observatories. SHWFSs were originally developed to address a problem that was introduced by the U.S. air force in the late 1960s to improve the images taken from satellites orbiting the Earth. Aden Mienel~\cite{History} came up with the idea of enhancing satellite images by limiting the exposure time in a way that wavefront error caused by the atmospheric turbulence would not exceed $\lambda/10$. This is similar to taking snapshots rather than long exposure times in order to prevent the accumulation of the blurring effect of the atmosphere. He also used measurements of the Optical Transfer Function (OTF)~\cite{GarciaRodriguez1978} of the atmosphere at the same time of the exposure to use it in post-processing in order to improve the sharpness of acquired images. As an astronomer, he was familiar with the standard Hartmann test, which he used to capture a snapshot of atmospheric aberration. At this point, several other scientists including Ronald Shack, joined the project. The experimental setup that the team was working on finally evolved to the first Shack-Hartmann sensor. The term "Shack-Hartmann" was first coined in 1984 by Ray Wilson at the Optical Sciences Center, University of Arizona~\cite{History}.

The SHWFS is currently the most common type of wavefront sensor used in AO systems. In addition to their role in AO systems, SHWFSs can also be used to measure the optical quality of a system in a way similar to that of interferometers~\cite{wfss}. Physical dimensions, sensitivity to vibration, and broader dynamic range are a few of the reasons why SHWFSs might be chosen over interferometric measurements for some applications.

\begin{figure}
\includegraphics[width=\linewidth]{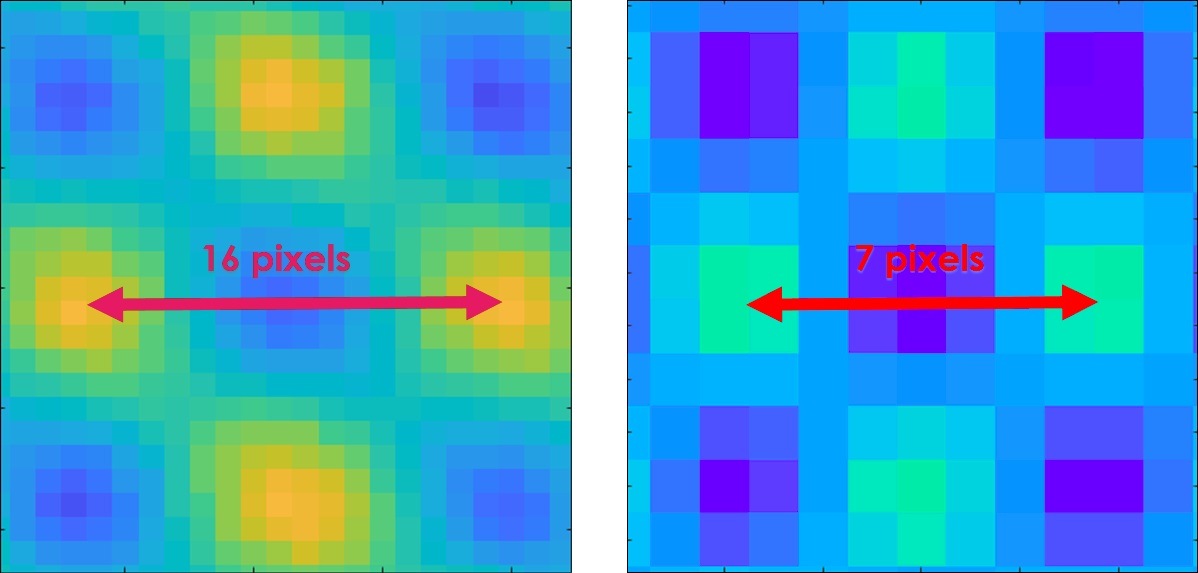}
\caption{Small crop of simultaneous WFS measurements of a waffle pattern on the Cilas prototype DM with HASO-128 on the right, and SSH-WFS on the left panel. Note the difference in resolving power between the two measurements. The red double arrow line indicates the distance equal to two actuator pitches on the DM surface.}
\label{ftssh1}
\end{figure}

The TMT has contracted with Cilas~\cite{boyer2018adaptive} to build the DMs for NFIRAOS. As part of that work, Cilas provided a prototype DM for testing. There are 616 actuators in the prototype DM (28 actuators across the diameter) while the NFIRAOS DMs will have 60 and 75 actuators across their diameters on a square grid~\cite{nfiraos}. A sampling resolution of at least 4 is sufficient to study high-frequency artifacts between the actuators. Figure~\ref{ftssh1} shows the difference in spatial resolving power between HASO-128, the highest resolution WFS commercially available at NRC's disposal, and SSH-WFS measurements. This would translate to a SHWFS with 252 for DM0 and 300 for DM 11 sub-apertures/lenslets. Such a SHWFS is not available commercially, which motivated the in-house design of our Super SHWFS~(SSH-WFS). In addition to the unprecedented resolution of this sensor, we considered cost as an important design factor, resulting in the fabrication cost of the final device to be less than 30k USD. This device is also able to record the wavefront at 50 frames per second at full resolution and up to 4 kHz for cropped regions of interest, making it possible to study the dynamics of the DM actuators. I also designed and performed verification tests to ensure that the SSH-WFS performance met its requirements.

In this paper, we present the design considerations and fabrication of the SSH-WFS device using low-cost solutions and commercial off-the-shelf components. We also present the opto-mechanical design as well as the results of differential tests and surface measurements down to 5~nm RMS error with this sensor.

\section{Design Outline}

\begin{figure}
\includegraphics[width=\linewidth]{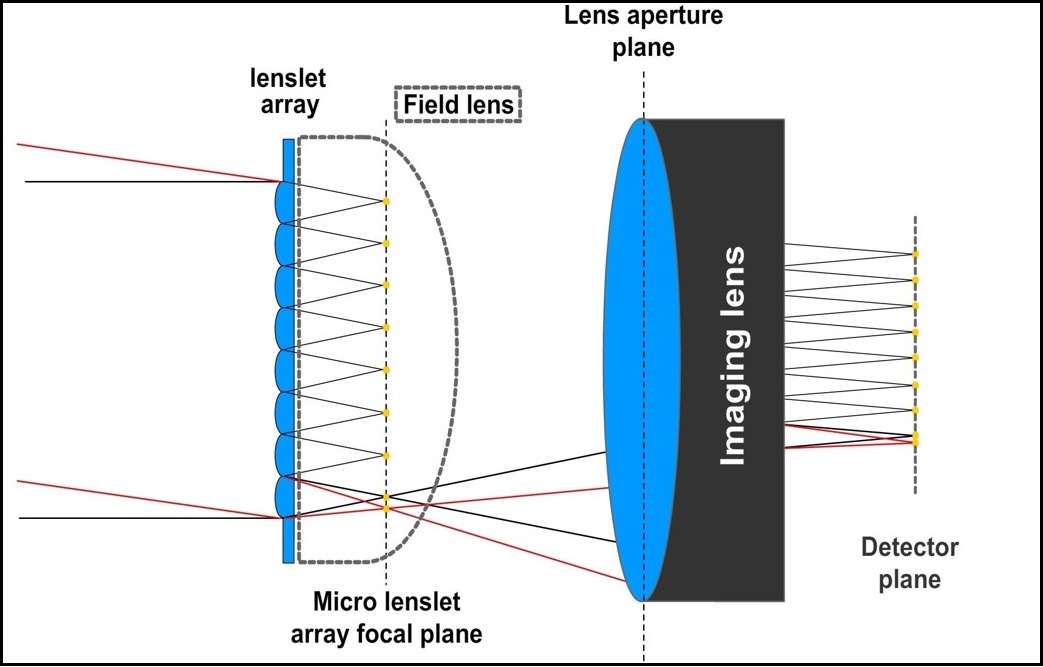}
\caption{The schematic of the design. The system consists of (from left two right): the lenslet array, field lens, imager lens which provides 0.69 magnification, and the detector.}
\label{14s}
\end{figure}

Figure~\ref{14s} shows a schematic of the optical design of this device. This system consists of:
\begin{itemize}
\item The detector/imager
\item The lenslet array
\item The optical relay/imaging lens
\item The field lens
\end{itemize}
Below is more information about the design and function of each main component.

The Super SHWFS incorporated the Andor Zyla 4.2 sCMOS camera. This camera provides 2048x2048 pixels array and full-frame reading frequency of up to 50 frames per second. We used multiple exposures for creating each single wavefront measurement to increase wavefront precision. This makes the high FPS and high quantum efficiency of this detector very desirable as it decreases the total duration of data acquisition. The read noise and linearity of the detector are respectively 0.9 e- and 99.8\% which also makes it a suitable option for this application.

\begin{figure}
\includegraphics[width=0.4\linewidth]{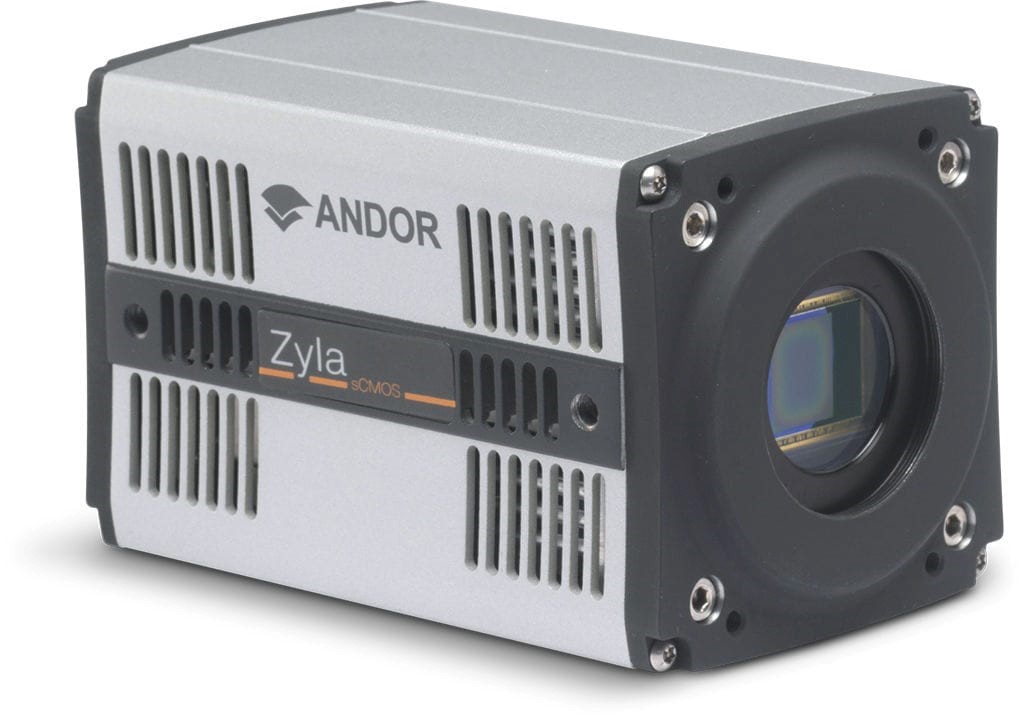}
\hfill
\includegraphics[width=0.4\linewidth]{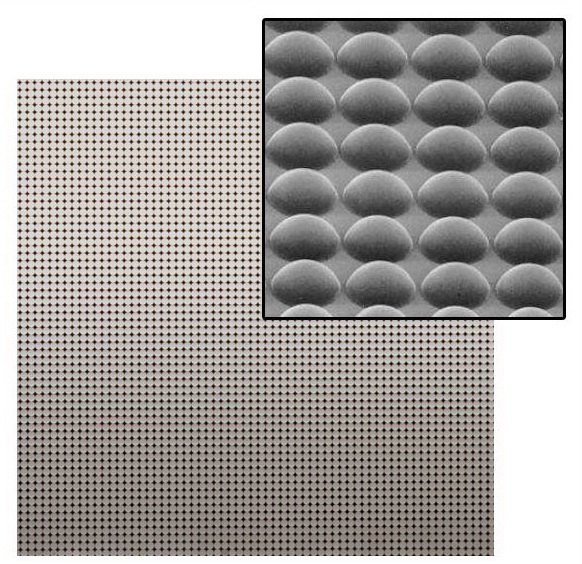}
\caption{The Zyla 4.2 camera and the schematic of the selected lenslet array.}
\label{s234}
\end{figure}

For the lenslet, we used a SUSS MicroOptics 75~$\mu m$ pitch, circular sub-aperture lenslet array (see Figure~\ref{s234}). Based on the design requirement, each sub-aperture (75~$\mu m$) spanned 8 pixels ($8\times6.5=52~\mu m$) which means an optical relay with the projection magnification of 0.69 is necessary between the lenslet and the imaging detector. More information about the camera and the lenslet array is presented in Table~\ref{t1s}.

\begin{table}
\centering
\caption{Specification of detector and lenslet array for SSHWFS.}
\begin{tabular}{l r l r }
\multicolumn{2}{c}{Detector specification} & \multicolumn{2}{c}{Lenslet array specification} \\
\hline
Active pixels & 2048 $\times$ 2048 & Dimension & 23.1 $\times$ 23.1 \\ 
Sensor size & 13.3 $\times$ 13.3 mm & Grid type & Square\\ 
Pixel size & 6.5 um $\times$ 13.3 mm & Aperture type & circular\\ 
Read noise & 0.9 e- & Pitch & 75 um\\
Dark current & 0.14 e-/pixel/sec & Focal length & 1.18 mm\\
Max QE & 82\% & Focal ratio & 15.73\\
Max dynamic range & 33000:1 & Refraction index & 1.458\\
Max frame rate & 53 fps & Thickness & 1.2 mm\\
\hline
\end{tabular}
\label{t1s}
\end{table}

\subsection{Role of the Optical Relay}
An optical magnification of 0.69 is required to project a lenslet to 8 pixels of the detector. We considered designing and fabricating this optical relay, but first considered commercially available, high-quality photography lenses to find a potential match for this application. The main criteria for choosing the right imaging relay are:

\begin{itemize}
 	\item  Equipped with an aperture larger than the cross section of the diverging beam (passing through the lenslet array), at the distance corresponding to the linear magnification equal to 0.69 times.
    \item  Provide a circular flat focal plane with the radius of 18.8~mm.
    \item  Reasonable low field distortion across the field to maintain linearity of phase measurements, meeting the required wavefront measurement precision of less than 5~nm across the field.
\end{itemize}

\begin{figure}
\includegraphics[width=\linewidth]{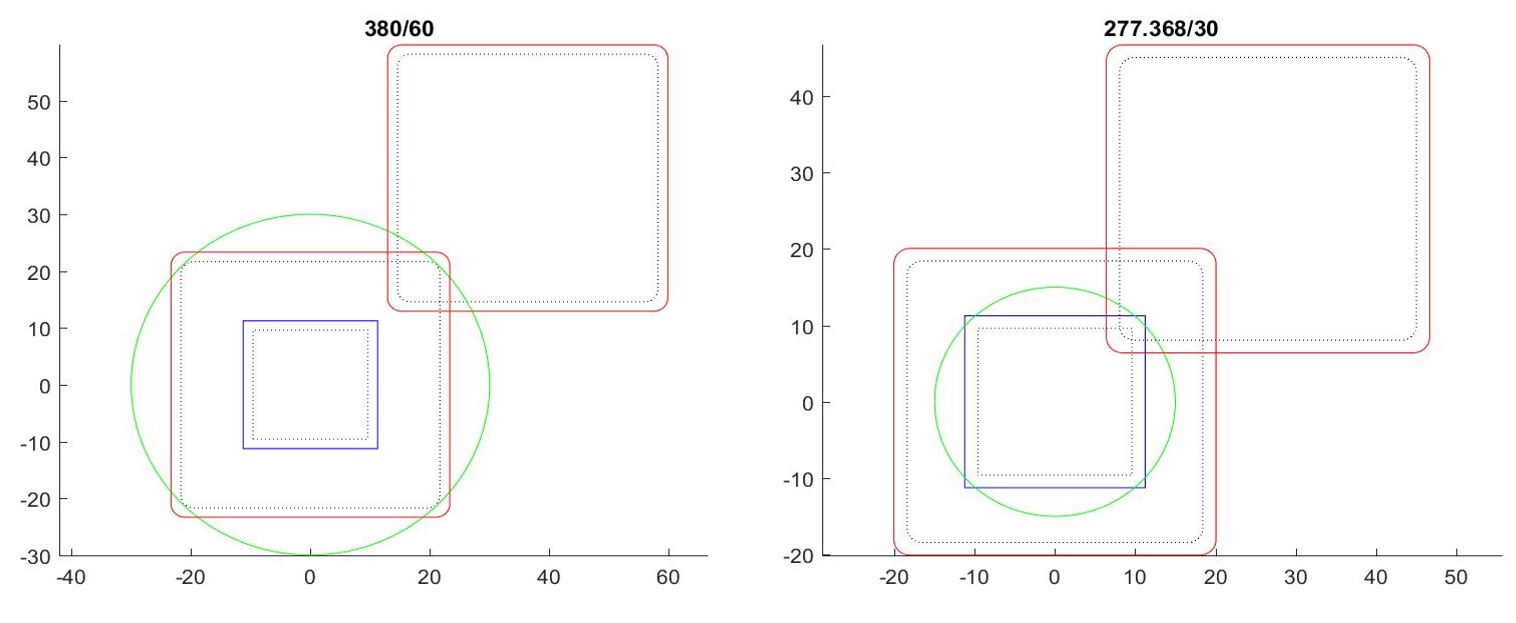}
\caption{Examples of simulation results for the two lenses. The blue rectangle shows the overall size of the lenslet array and the dashed blue square shows the areas of interest on the lenslet array (256 sub-apertures wide). The green circle represents the lens aperture and the red squares are the beam footprint on the lens aperture plane in zero and maximum tilt/tipped condition. The blue squares are on the plane of the lenslet array and the rest are on the plane of the lens aperture. Although the chosen lens can fit the zero field beam (central red square) in its aperture (the left panel), none of the lenses in the survey were able to contain the tilt/tipped beam footprint without additional considerations in the design. We resolved this issue by utilizing a field lens (see Section~\ref{flens}) right after the lenslet array position.}
\label{15s}
\end{figure}

\begin{figure}[htbp]
\centering
\includegraphics[width=0.25\linewidth]{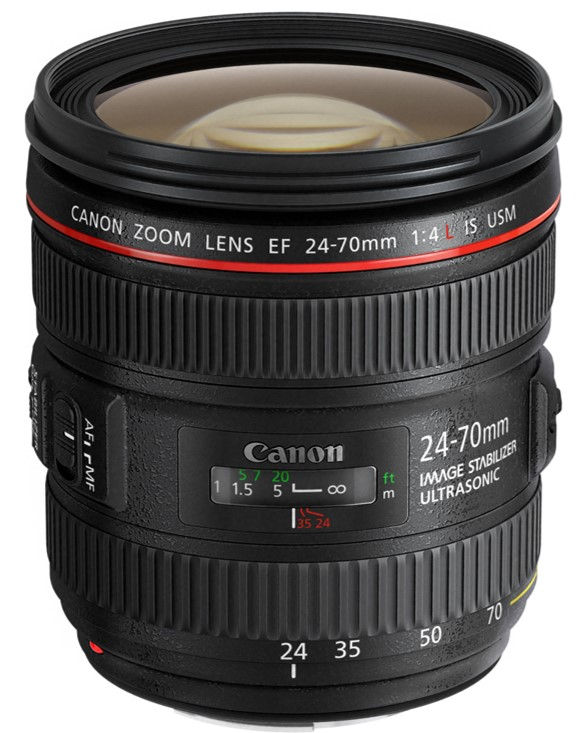}
\includegraphics[width=0.4\linewidth]{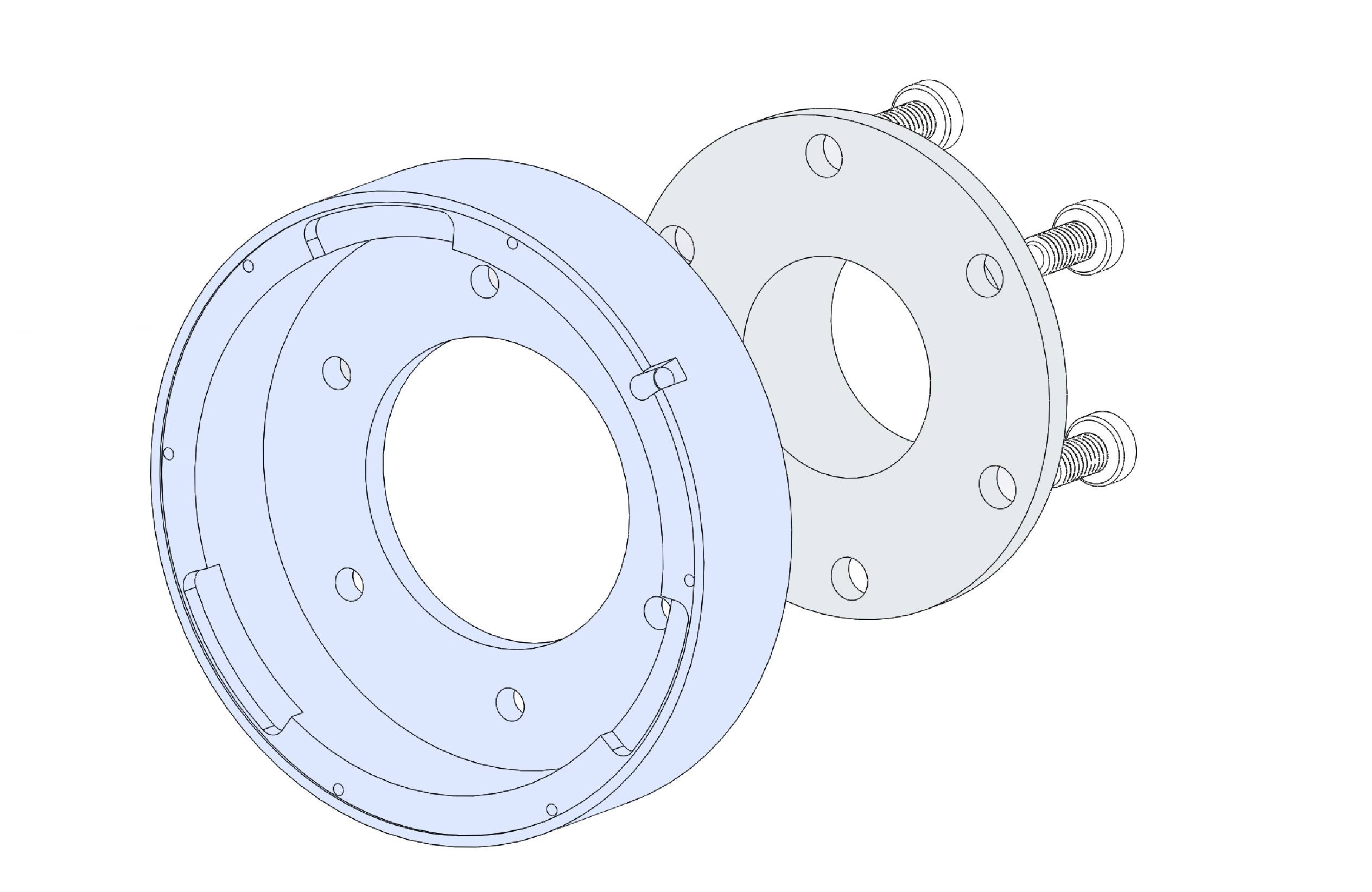}
\includegraphics[width=0.3\linewidth]{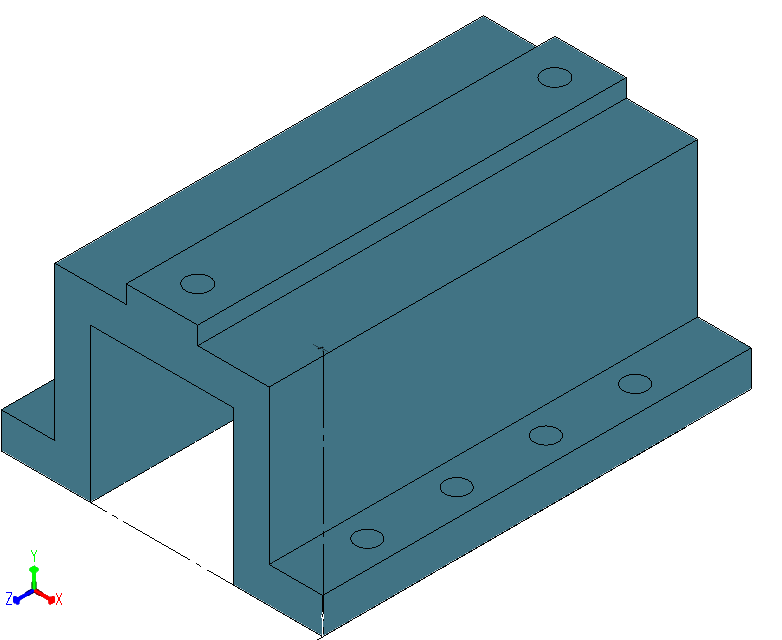}
\caption{Left: The Canon EF 24-70mm f/4 lens chosen for the design of SSH-WFS. Middle: The CAD design of the adaptor, which mounts Canon lens on the Zyla camera. Right: The CAD design of the mount which places the Zyla camera on the optical plate of SSH-WFS.}
\label{fClens}
\end{figure}

The first option to provide such an optical relay, is to look at commercially available options in order to keep costs as low as possible. We gathered information for 35 commercially available lenses and inputted their characteristics into a simulation. The said simulation takes the projected size of the lenslet array illumination pattern at the distance to the lenses that corresponds to the magnification of 0.69 for the center of the field and compares this with the cross section to the lens input pupil size. The result of this study determined that the Canon EF 24-70mm f/4 lens (see Figure~\ref{fClens}) is the best choice for this application. The selected lens is able to contain the footprint of the zero-tilted beam passing through the lenslet array and projects it to the detector at the suitable magnification. However, as Figure~\ref{15s} shows, there are still challenges with tilted rays. We address this issue in the next section, by introducing a field lens (see Section~\ref{flens}) to the design. The combination of using a field lens and off-the-shelf options for the optical relay significantly decreased the overall cost of the project.

To position the Canon lens on the Zyla camera and to install the system on the optical bench, we also designed the lens adaptor and mounting structures. These designs were fabricated at the NRC machine shop facilities and installed on the bench. The 3D CAD design of these parts can be seen in Figure~\ref{fClens}.

\subsection{The Field Lens: Low-Cost Solution for the Field Illumination Challenge} \label{flens}
\begin{figure}
\centering
\includegraphics[width=0.30\linewidth]{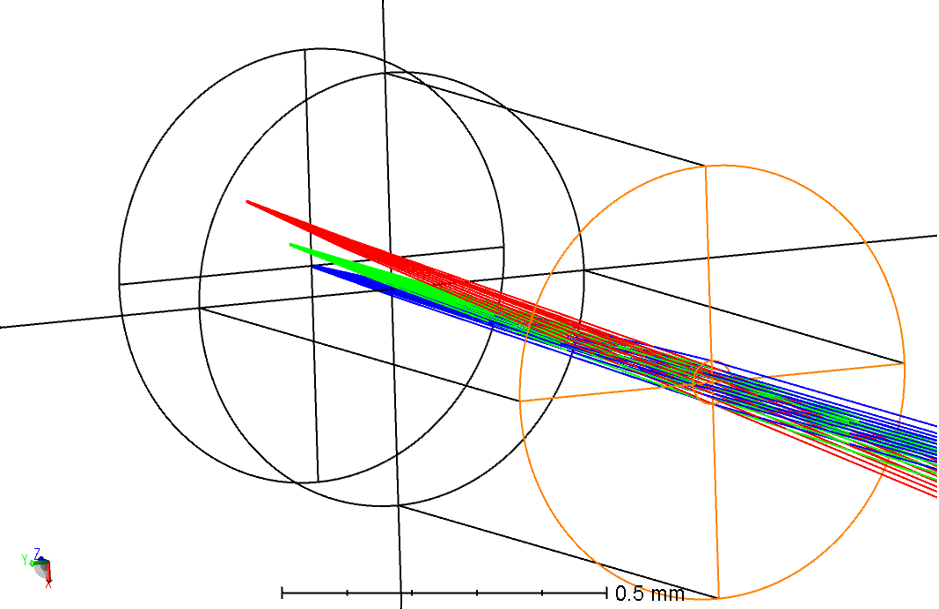}
\includegraphics[width=0.20\linewidth]{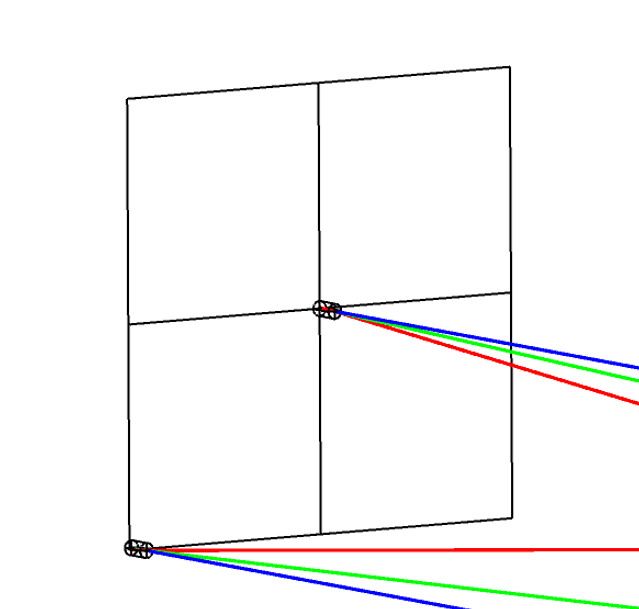}
\includegraphics[width=0.40\linewidth]{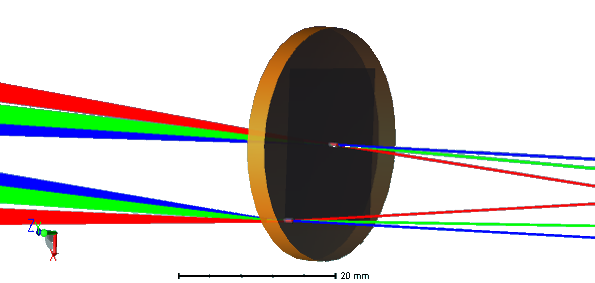}
\caption{Left: The Zemax model we provide for a single lenslet based on the SUSS MicroOptics specifications. Middle: The lenslet array model with a lenslet on the center and one on the farthest lenslet from the center. Right: The Zemax model of the lenslet array in the presence of the Field lens (orange transparent object). The different ray colour represents a range from minimum to the maximum tilt of the incoming beam. Combining maximum tilt with the position of the lenslet right on the corner of the array, represents the worst case illumination scenario that the field lens design should be able to handle.}
\label{fshz}
\end{figure}

Although the chosen lens pupil is large enough to contain the beam footprint created by the lenslet array, it is unable to catch the incoming beam in the maximum tilt/tipped angle dictated by the requirement. To resolve this issue, we found an inexpensive and efficient solution which is to add a field lens immediately after the lenslet array. This plano-convex lens causes tilted beams to converge into the aperture of the lens without having major effects on the other optical properties of the system. To estimate the specification for the field lens, we provide a Zemax model of the lenslet array and analysed the performance of the system for a range of tilt angles of the incoming rays (see Figure~\ref{fshz}).

\begin{figure}
\includegraphics[width=1\linewidth]{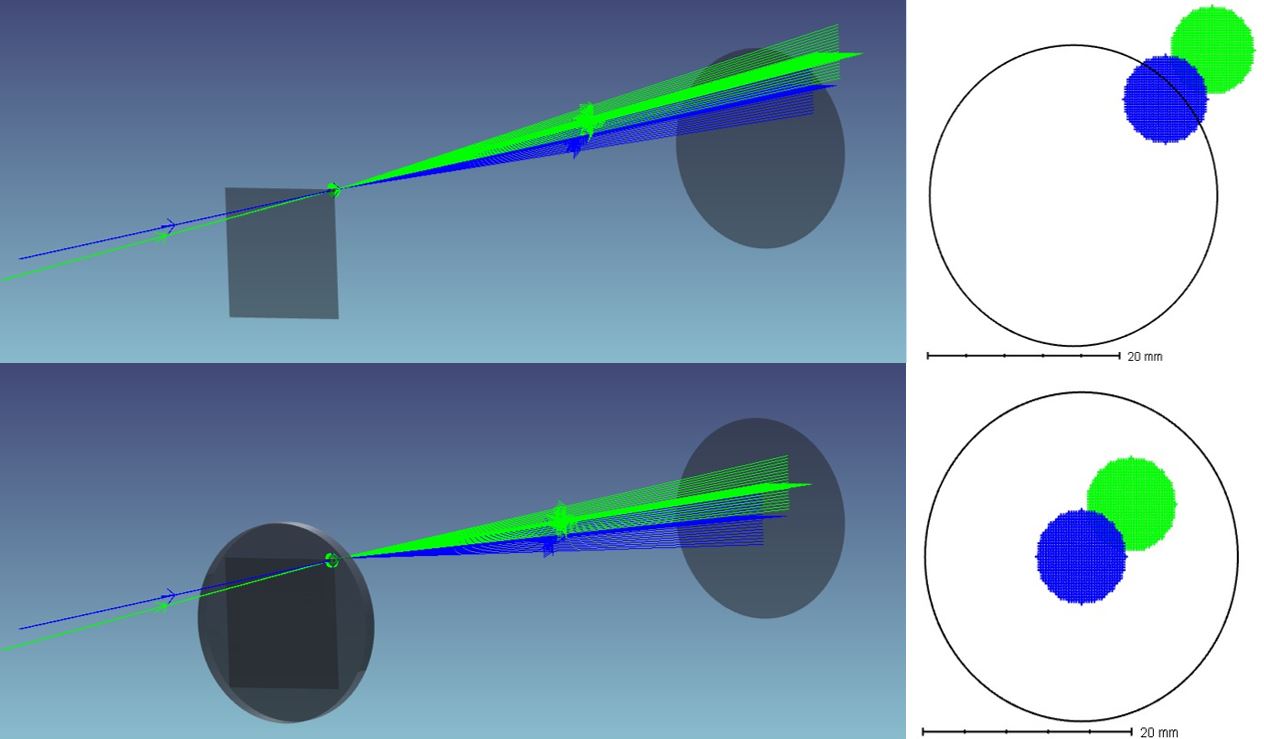}
\caption{Zemax designs showing the effect of the field lens for a sub-aperture on the corner sub-aperture for the zero and maximum tilt/tipped beam (blue and green respectively). The green scenario represents the most tilted and shifted beam considered in the requirements relative to the zero field central beam. The beam profile on the lens aperture plane could be seen on the right. Top: No field lens: the tilted beam would fall outside the lens aperture so the sensor would not be able to sense the wavefront on some sub-pictures in the tilted situation, as well as miss some light in the zero tilt scenario for the edge sub-apertures. Bottom: using a field lens immediately after the lenslet array: even the footprint of the tilted beam hitting the lenslets on the corner of the array would be captured by the imaging lens entrance pupil.}
\label{16s}
\end{figure}

Figure~\ref{16s} shows the effect of the field lens using the system model designed in the Zemax environment. By incorporating a proper field lens into the design, the aperture of the optical relay could be smaller by a factor of 3.5 and still gather the tilted beam footprint. This dramatically decreases the cost and complexity of the system. 

\begin{figure}
\centering
\includegraphics[width=0.35\linewidth]{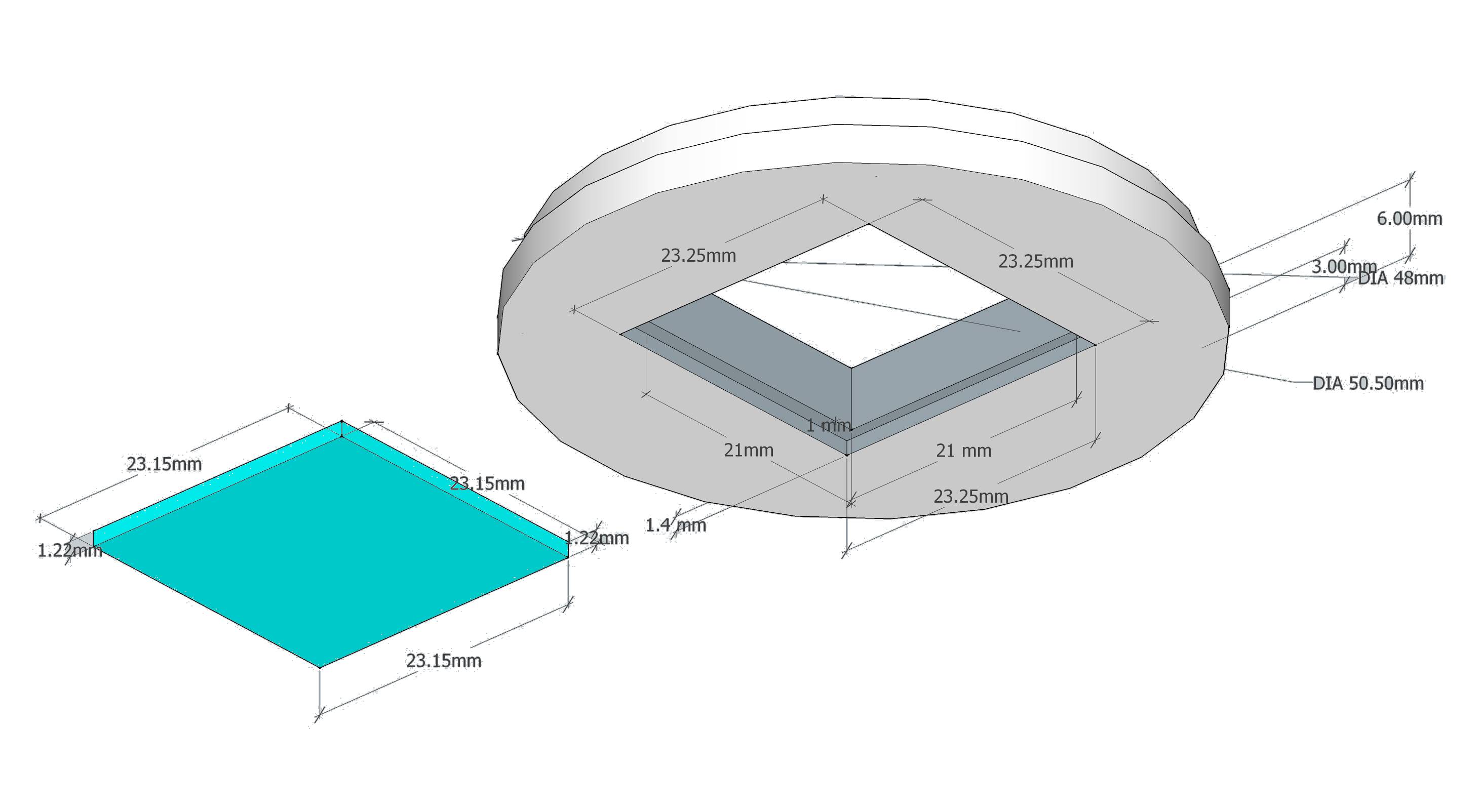}
\includegraphics[width=0.25\linewidth]{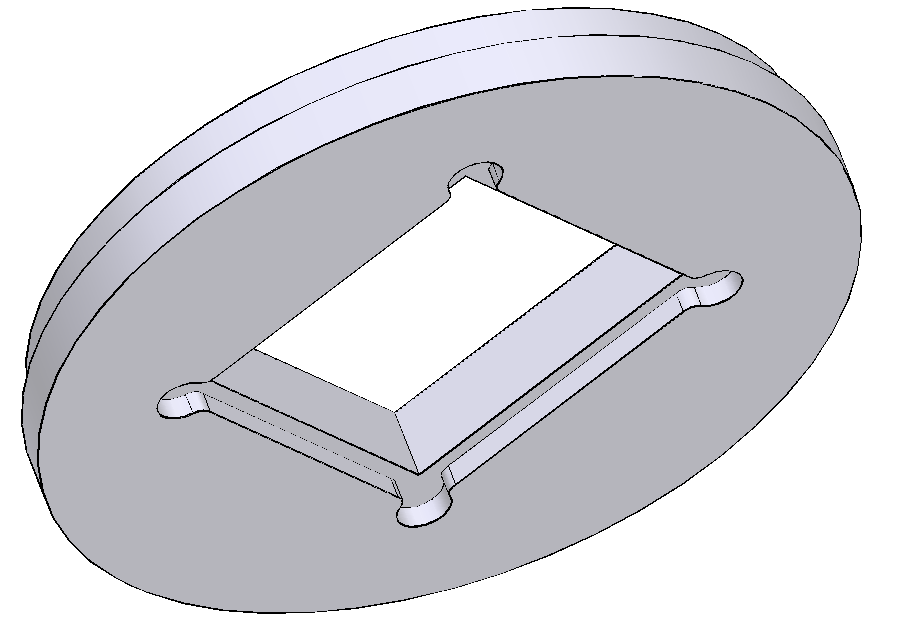}
\includegraphics[width=0.25\linewidth]{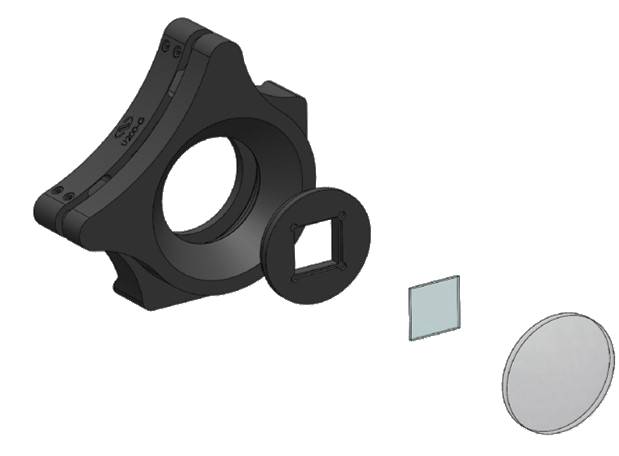}
\caption{Left: The CAD design of the mechanical part that fixes the lenslet to the flat part of the field lens. The blue box represents the mechanical model for the lenslet array. Middle: The final design of the same part before fabrication. Note the round edges on the corner that were added to simplify the machining process. Right: The CAD model of the whole assembly. The flat side of the lenslet array is outward and the field lens inward relative to the page. The custom-made part holds the lenslet array inside the rectangular edge, then presses it to the flat side of the field lens. The whole assembly then screws on the tilt/tip mount which secures the edge of the curved side of the field lens against the flat area of the custom made piece.}
\label{flh}
\end{figure}

In addition to the optical design, a practical mechanical design was also needed to correctly hold the field lens and the lenslet array in place. The lenslet array is extremely fragile and sensitive to touch, so therefore careful consideration was necessary to design a mechanism to keep it in place. We intentionally chose the field lens to be plano-convex, so the flat side of the lenslet array could be fixed on the flat side of the field lens. Both optical elements are then fixed by a custom designed mechanical piece to a tilt/tip adjustable mirror mount. Such a design ensures the safety of the featured side of the lenselet array, as well as places both optical elements as close as possible to the pupil conjugation plane which is essential for the correct performance of the SSH-WFS. Figure~\ref{flh} shows the mechanical design of the lenslet array-field lens installation.

\section{Image Reduction and Slope Measurement}
\begin{figure}[htbp]
\centering
\includegraphics[width=0.8\linewidth]{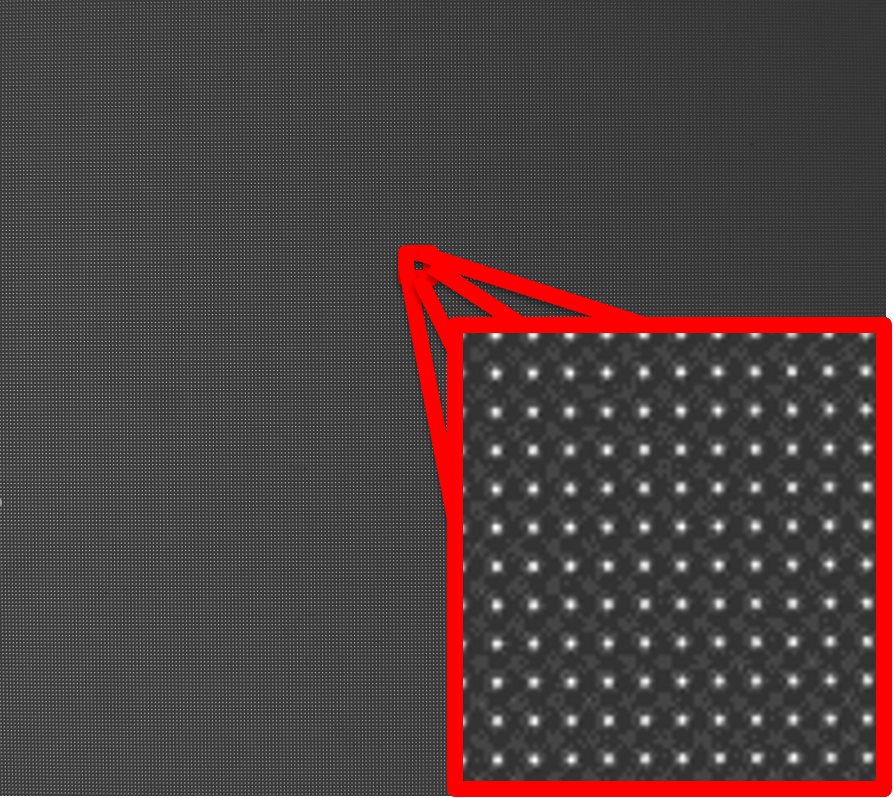}
\caption{A raw image of the Zyla detector while measuring wavefront. The grid of dots is created in the focal plane of the lenslet array. A small region is shown magnified.}
\label{fdet}
\end{figure}

We also developed a code package exclusively for the SSH-WFS which reads the WFS detector, applies necessary considerations (i.e applying pupil masks, removes illumination gradient, etc.), and calculates the measured wavefront. A view of what the Zyla detector measures can be seen in Figure~\ref{fdet}.

The precision of the final wavefront is dependent on the process that determines each sub-aperture spot position. Therefore, we developed an advanced multi-stage image reduction method. We use a zonal, 32x32 segment histogram thresholding algorithm as the phase-one reduction process to roughly find the position of active sub-apertures and recognize the grid connecting them. This method is robust against intensity changes across the field of view, which is one of the main sources of centroiding error in such measurements. In phase two, a more sophisticated histogram thresholding applies specifically to each sub-aperture detected in phase one. The final position is calculated by the intensity-weighted average of validated pixels, which passes through the phase two thresholding process.

In the next step, the precise grid connecting all illuminated sub-apertures is constructed as well as the position of shadowed sub-apertures and the pupil edge are being determined. The code also measures the clocking of the lenslet array (in case it exists) and applies correction to the final positions before the wavefront reconstruction process. Each dataset contains multiple exposures to increase the precision of the spot positions. To reach the $5~nm$ RMS wavefront residual requirement across the pupil, typically 10-25 exposures per dataset are taken. This means each data acquisition session takes up to 1.2 seconds on the detector in the low-noise mode, which is less than the 5 second requirement of the wavefront measurement. 

\begin{figure}[htbp]
\centering
\includegraphics[width=0.8\linewidth]{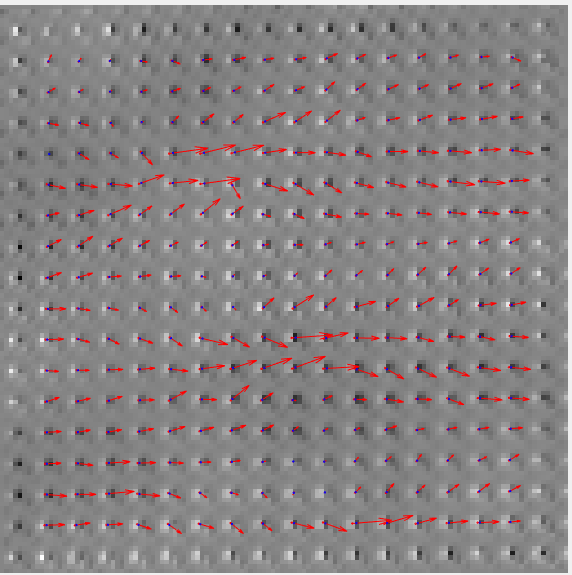}
\caption{This plot shows the difference between a reference and a non-flat measurement for a small region of the detector. The wavefront turbulence is injected into the experiment by placing a transparent plastic CD cover into the pupil conjugation. The measured vectors on the top represent the slope measured on each sub-aperture. The difference between the two spots in the presence of no turbulence should ideally be zero. However, the motion of spots due to the injected wavefront turbulence causes the spots to move. The difference of two spot PSF which are slightly displaced is a sine wave like artifact. Note that a more significant difference between the two PSF shaped spots corresponds to a larger slope measurement.}
\label{fmes1}
\end{figure}

To measure one set of relative wavefront slopes, the process described above is applied on each exposure of the reference and measurement datasets. The results are two sets of precise measurements of the spot positions for each sub-aperture. A 3-sigma clipping and center-of-gravity algorithm is used to reach the high-precision position of the spot in each sub-aperture. After calculating the spot position map, the effect of potential dust particles is removed by interpolating all the inactive sup-apertures which have more than 3 valid neighbors. Repeating the procedure above provides a set of high-precision measurements of relative movements of spots for all sub-apertures across the field. Using this information and the known physical scale of the system, it is possible to calculate X and Y derivatives of the unknown wavefront relative to the reference measurements and eventually reconstruct the phase wavefront. Figure~\ref{fmes1} shows the representation of spot displacement between the two measurements (reference and actual measurement) by plotting the difference between the two measurements in addition to the measured slopes. 

\section{Wavefront Reconstruction}
Solutions based on linear decomposition and matrix inversion are the simplest and most common approaches for wavefront reconstruction, however they require large buffer memory for calculations. The number of minimum required memory bytes is usually of the order of [number of sub-apertures]$^{2}$ which in this case is $2^{32}$ bytes. The actual number is even larger than this estimate considering the necessary bytes for storing float precision and other overheads. To make it possible to run our wavefront re-constructor on a typical commercial computer, we used a more complicated approach of iterative wavefront re-construction. We developed an iterative wavefront reconstructor based on the method described by W. Southwell~\cite{southwell}. This method is especially beneficial for reconstructing wavefronts measured from a large number of sub-apertures. This method uses two order of magnitude less memory compared to methods that are based on direct matrix inversion. The iterative core argument is:

 $$\phi^{(m+1)}_{jk}=\phi^{(m)}_{jk}+ \omega [ \overline{\phi} ^{(m)}_{jk}+b_{jk}/g_{jk}-\phi ^{(m)}_{jk}]$$
 
\noindent where $\phi$ is the phase, $j$ and $k$ are position indices and $m$ is the iteration counter. Also:
 
$$ b_{jk}=[S^y_{j,k-1}-S^y_{j,k}+S^x_{j-1,k}-S^x_{j,k}]h$$
$$\overline{\phi}_{jk} =[\phi_{j+1,k}+\phi_{j-1,k}+\phi_{j,k+1}+\phi_{j,k-1}]/g_{jk}$$
$$\omega =\frac{2}{1+sin[ \pi /(N+1)]} $$
$$g_{jk}=\begin{cases}2 & j =1  \vee  N;k=1  \vee  N \\3 & \begin{cases}j =1  \vee  N;k=2\ to\  N-1\\k =1  \vee  N;j=2\ to\ N-1\end{cases} \\4 & otherwise\end{cases} $$

\noindent where $h$ is the grid pitch, $N$ is the number of sub-apertures across the lenslet array, and $S^x_{jk}$ is the x-slope in the position $j$ and $k$. The full description and calculation of the method can be found in Southwell's paper~\cite{southwell}.
 
We also developed a fast Zernike wavefront reconstructor for reconstructing the wavefront from the measured slope matrix. The slope matrix $\begin{bmatrix} X_{slope}\\  
Y_{slope}  \end{bmatrix}$ can be written as $Zs \times C$, where \textit{Zs} is the Zernike slope reconstruction matrix and \textit{C} is the coefficients vector. For creating matrix \textit{Zs}, the X and Y derivation of the first 30 radial orders are calculated and rearranged to the vector form. Each vector becomes a column in matrix \textit{Zs}. The result is a matrix of size [number of valid sub-apertures $\times$ 2, number of reconstructing Zernike modes]. By calculating Zs and measuring $\begin{bmatrix} X_{slope}\\ 
Y_{slope}\end{bmatrix}$, it is now possible to calculate C using SVD or similar inverse matrix calculation methods. The result, Vector C, contains intensity coefficients for each Zernike mode. The next step is to create matrix \textit{Z}, calculated by the exact same process as matrix \textit{Zs}, except that instead of derivations of Zernike polynomials, it consists of the polynomials themselves. Finally, the reconstructed phase wavefront can be then written as $PWF_{recons.}=Z \times C$.

\section{Verification Tests}
\begin{figure}[htbp]
\centering
\includegraphics[width=0.459\linewidth]{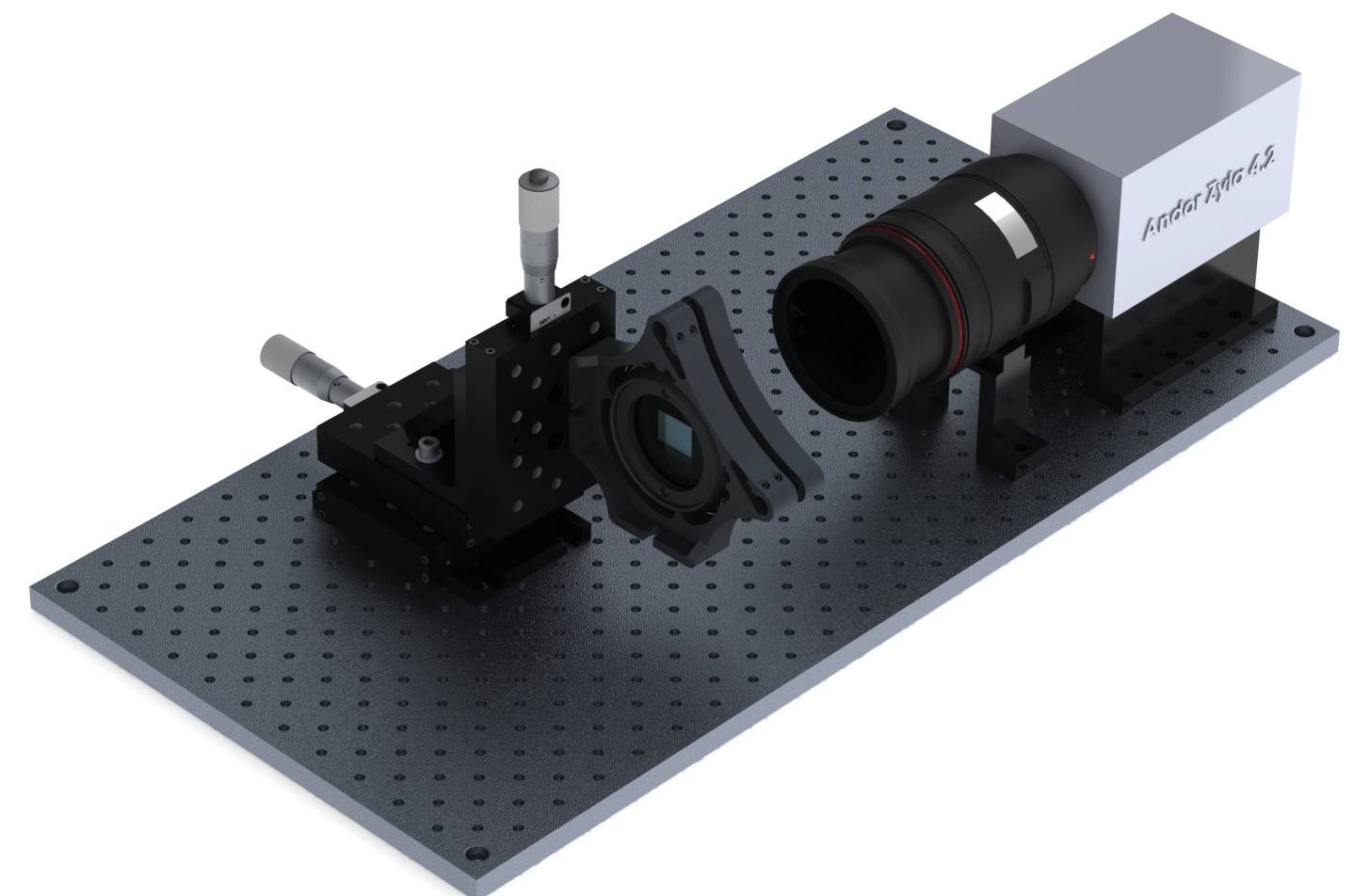}
\includegraphics[width=0.459\linewidth]{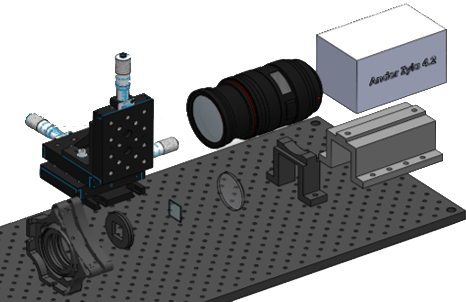}\\
\includegraphics[width=0.9\linewidth]{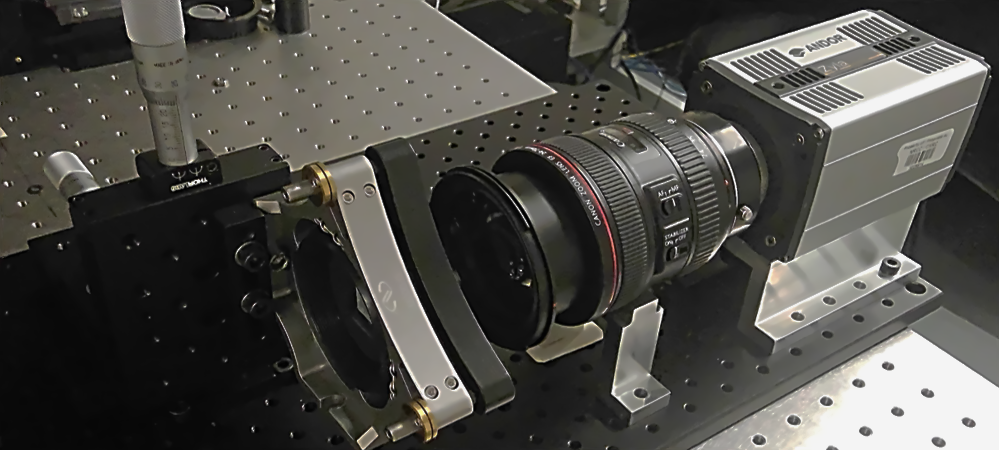}
\caption{Top: The initial CAD design of the device and the exploded view of all components. The CAD design employs precise optical calculations made by the Zemax model of the device. Bottom: Actual fabricated SSH-WFS device.}
\label{ffd}
\end{figure}

Figure~\ref{ffd} shows the CAD of the SSH-WFS device as well of the actual device. All parts are installed on a separate optical breadboard to ease moving the device from a bench to another for different experiments and applications.

\begin{figure}[htbp]
\centering
    \includegraphics[width=01\textwidth]{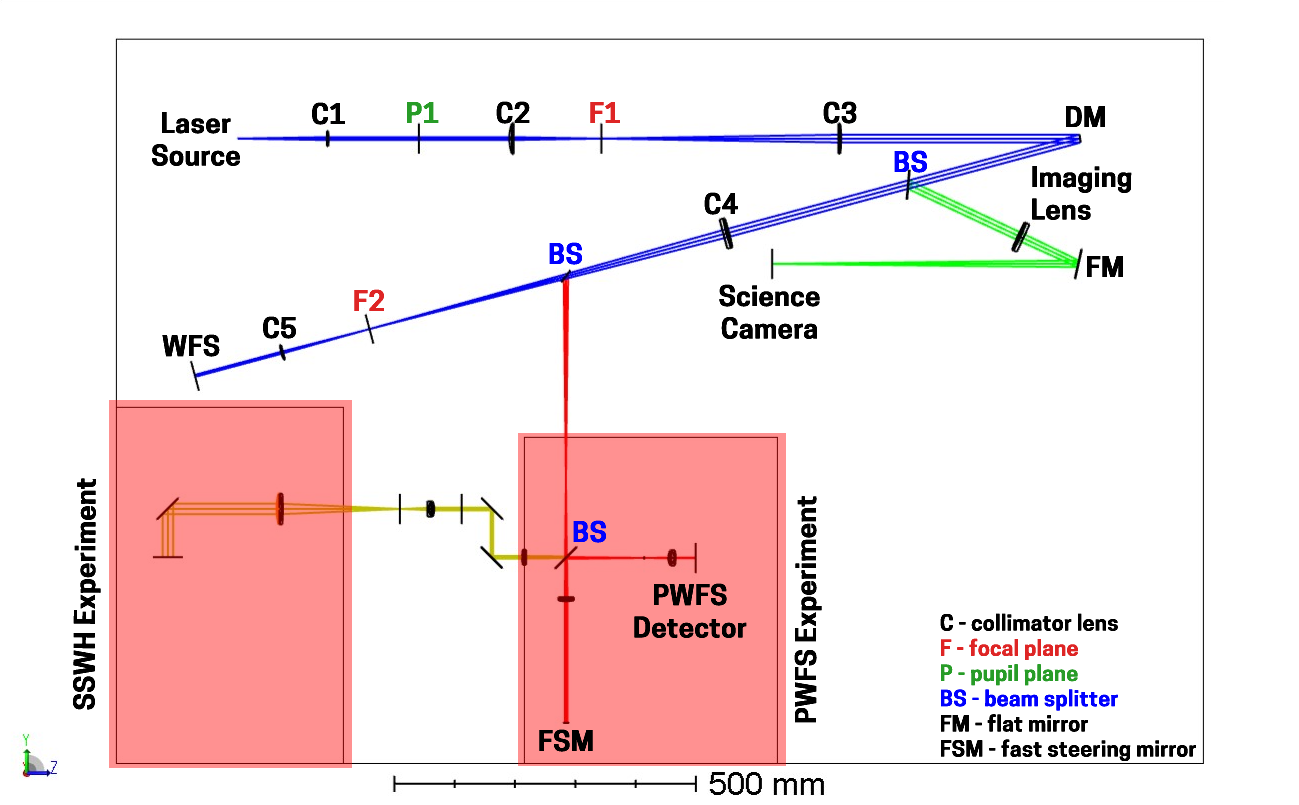}
\caption{Schematic of the test bench used for series of test on SSH-WFS.}
\label{f32s}
\end{figure}

Before using the SSH-WFS for its ultimate application, it is necessary to characterise and verify its precision and noise propagation behaviour. We designed a series of tests to verify the device performance and confirm that it meets the measurement requirements. The test bench can be seen in Figure~\ref{f32s}. 

On this bench, we used the combination of an Alpao DM 97-15 and HASO 4 broadband SHWFS simultaneously with the SSH-WFS to perform these test scenarios. In some tests, we also used the actual Cilas DM prototype and much higher-order (as compared to HASO4) SHWFS, HASO 128 to compare the abilities of the SSH-WFS to the highest resolution commercial SHWFS at our disposal. The closed-loop design of the test setup gave us the ability of injecting custom wavefronts in a controlled loop scheme.

\subsection{Performance comparison}

\begin{figure}[htbp]
\centering
\includegraphics[width=0.8\linewidth]{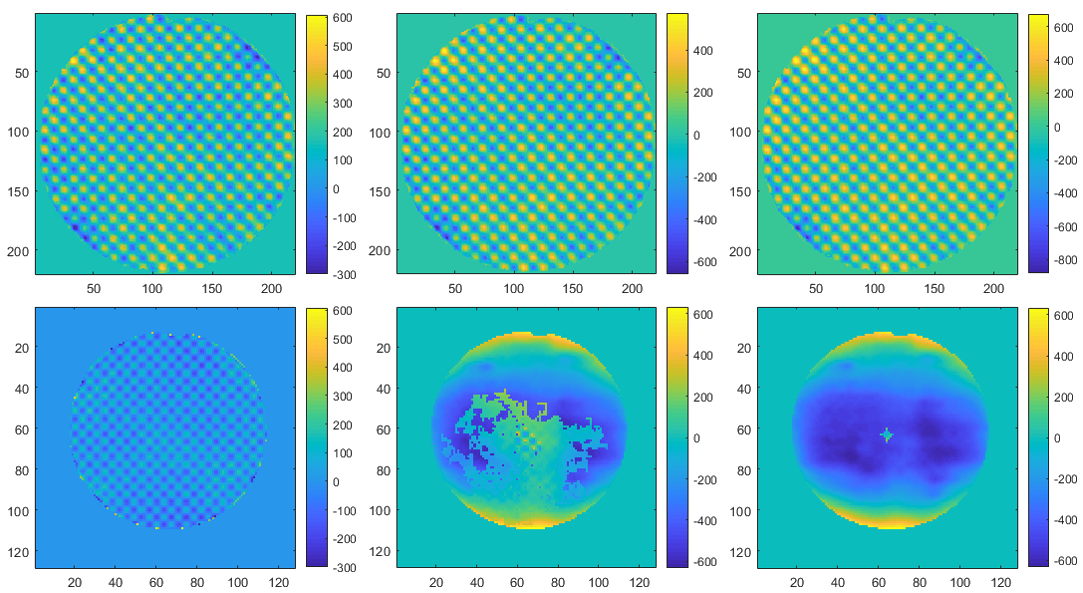}
\caption{Measurement of the waffle pattern on the Cilas DM prototype by SSH-WFS (top) and HASO~128 (bottom). The actuator push increased from left to right by 25\%, 50\% and 75\% of the maximum actuator stroke. Both WFSs are performing similarly on a 25\% stroke. However, the HASO dynamic range is saturated in the middle and right panels which caused corrupted measurement.}
\label{forg1}
\end{figure}

\begin{figure}[htbp]
\centering
\includegraphics[width=0.8\linewidth]{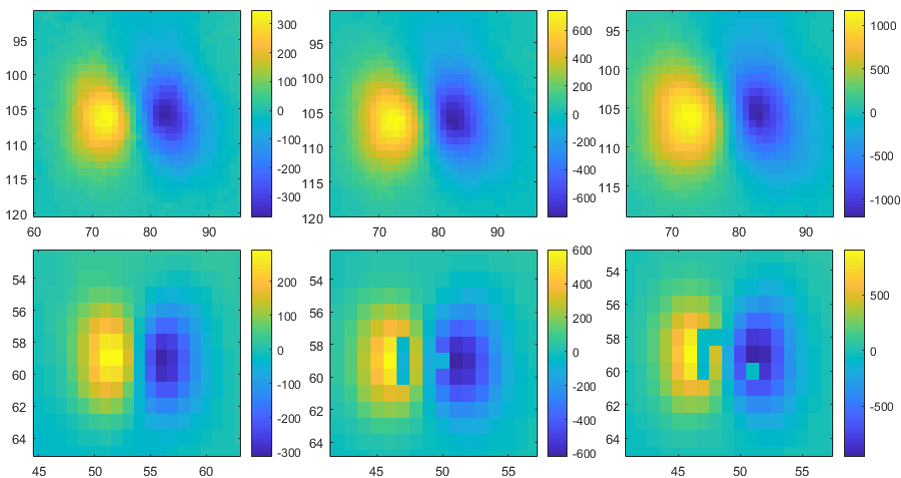}
\caption{Measurement of the single push-pull pattern on the Cilas DM prototype by SSH-WFS (top) and HASO~128 (bottom). Push-pull pattern consists of pushing and pulling a pair of neighbor actuators. The actuator push increased from left to right by 25\%, 50\% and 75\% of the maximum actuator stroke. Also, both WFSs are performing similarly on a 25\% stroke, the HASO dynamic range is saturated in the middle and right panels, which caused corrupt measurement. Additionally, the difference in resolving power between the two WFSs are well represented in this figure.}
\label{forg2}
\end{figure}

We performed a series of performance comparison tests to represent the advantages of the SSH-WFS over commercially available SHWFSs for certain applications. For these tests, we had the opportunity to use a prototype of the NFIRAOS DM produced by Cilas on a different test bench. The main relative advantages of the SSH-WFS are measurement resolution and dynamic range. Figure~\ref{forg1} and \ref{forg2} are representing a real case of such advantages for the SSH-WFS over the HASO~128 SHWFS.

\subsection{Stability Test}
\begin{figure}
\centering
\includegraphics[width=0.9\linewidth]{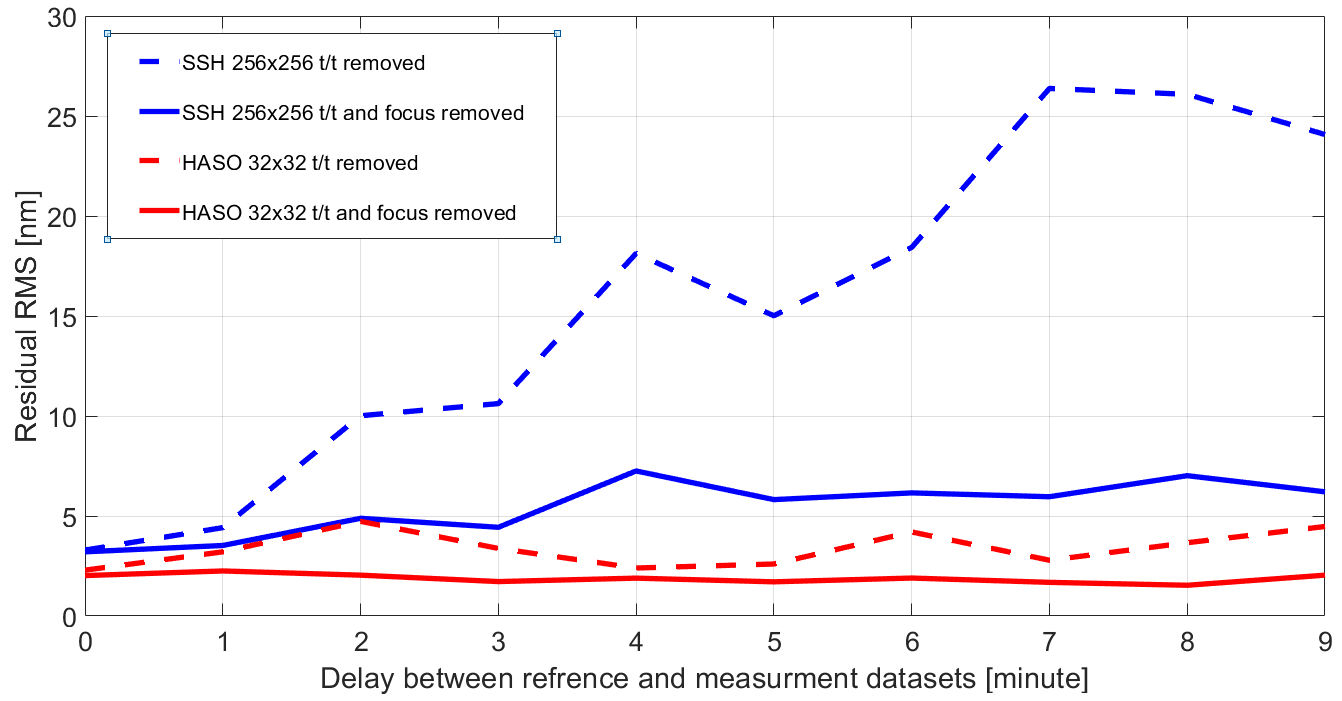}
\caption{The stability test result. Blue solid line describes the response of our 256$\times$256 Shack-Hartman sensor on the experimental bench. The solid line is for tilt/tip and focus removed performance, and the dashed line is for tilt/tip removed performance. The red lines are the same data for the 32$\times$32 HASO Shack-Hartmann device for means of comparison.}
\label{f36s}
\end{figure}

We also designed a series of tests to measure the stability of the system. These measurements determine how long a single wavefront reference is valid and how frequent the measurement of the reference should be repeated. The stability is determined by the thermal and mechanical properties of the test bench and its environment, and is not an intrinsic property of the device. For this test, we measured the wavefront residual of a constant wavefront with a delay of one minute up to nine minutes between the reference and measurement data sets. In an ideal situation, all residual measurements should be constant and around the nominal precision of the device. However, internal effects like the randomness of readout noise and environmental effects like the thermal instability of the bench would cause non-zero wavefront residual. It should be noted that we do not use any thermal stabilization solution on the current experimental bench and the temperature on the bench can vary in a range of 10~C. In the case of specific applications, the stability performance could be improved by implementing an athermalized design. Figure~\ref{f36s} shows the relation between the RMS residual and the delay between the actual and reference measurements. As expected, these measurements show that the focus mode is changing the most. This is to be expected since slight changes in temperature change the optical path length during measurements. However, the SSH-WFS device as mentioned previously, is mainly designed to study higher spatial frequency modes such as actuator dynamics and surface behaviour between actuators. Therefore, removing the lowest frequency modes such as tilt, tip and focus does not impact the device's ability to satisfy its design requirements. For a tilt, tip, and focus removed measurements (solid lines), the behavior of our system is favorably comparable to the certified calibrated system.

Combining these results with the precision result of sub-section~\ref{sshprec}, we show that this system is capable of providing a minimum of 180/60 (not considering focus applications/considering focus applications) seconds of stability between each reference measurement for the lab environment if 5nm RMS precision requirement is to be satisfied.

\subsection{Scale Calibration}
\begin{figure}[h]
\centering
\includegraphics[width=0.6\linewidth]{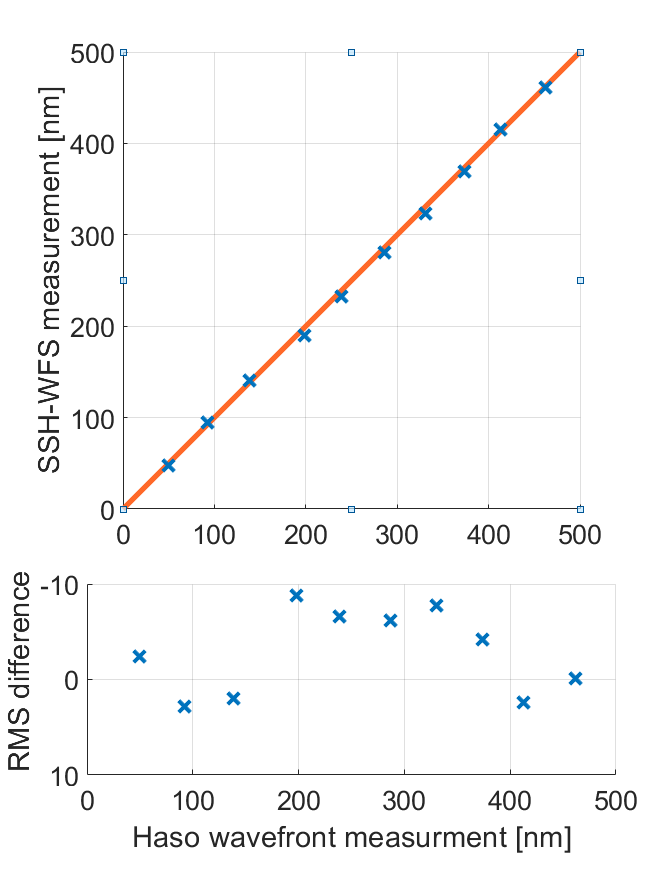}
\caption{The correlation between HASO and SSH-WFS measurement. This result shows the scale of SSH-WFS is well calibrated for a large range of wavefront input.}
\label{f34s}
\end{figure}

The scale of the SSH-WFS measurements is calculated based on the optical geometry and the lenslet specifications. We measured a series of differential patterns simultaneously using the HASO~4 and SSH-WFS to check the linearity and scale of the measurements. All measurements were relative to the second reference measurement to eliminate the effect of NCPA on different optical paths between the HASO and our wavefront sensor. Based on the manufacturer's specification, the repeatability of the HASO wavefront sensor in the range of experiment is 3~nm RMS. The pupil size for the SSH-WFS is a circle with a diameter of 256 sub-apertures and for the HASO it is the same physical size but with 32 sub-apertures across. The resolution of the two sensors are very different, so to avoid aliasing on the HASO WFS, we only used very low spatial frequency patterns for this test. Results are shown in Figure~\ref{f34s}. The red line represents $X = T$. The fit to the data has a correlation of 0.9996 and the RMS deviation from the fit is 4.2~nm, which shows the acceptable level of precision in the scale calibration.

\subsection{Noise characteristics}\label{sshprec}
The precision of the measurements are dependent mostly on two factors, the number of stacked exposures for each data set and the signal to noise ratio of the measurements. We performed a series of tests to characterize the noise floor based on these two factors. For these tests, three series of 25 consequent exposures (0.01 second each) of a constant wavefront were recorded for different light source intensities. The intensity of the light source was set to full well of the detector for the first series, and then was dimmed by a factor of 10x and 100x on the subsequent two sets of exposures. For this scenario, we utilized the Zyla detector in the low-noise, slow read (28 fps) and full well capacity (16-bit) mode to reduce the read noise as much as possible.

\begin{figure}[h]
\centering
\includegraphics[width=0.9\linewidth]{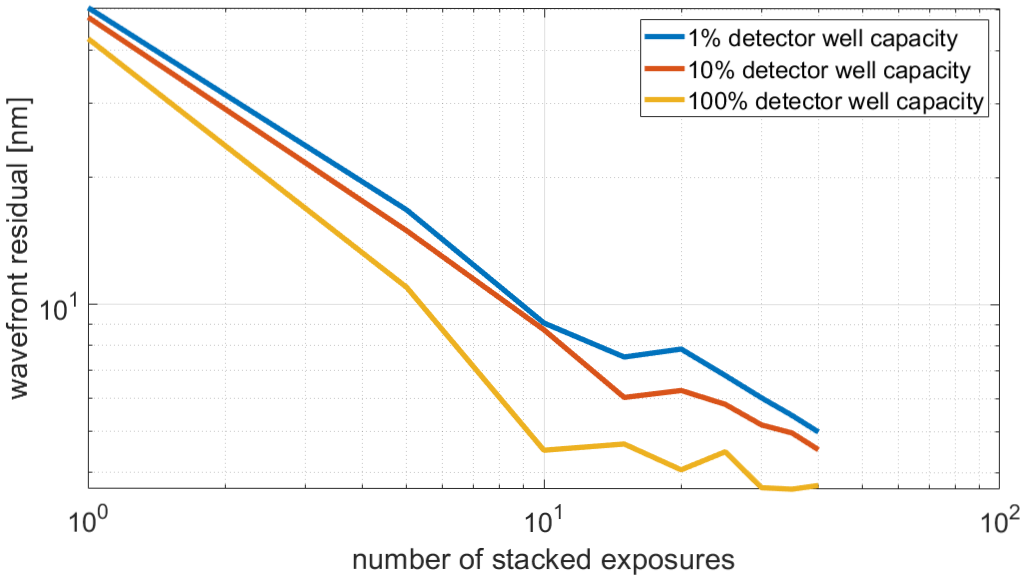}
\caption{The wavefront residual for two sets of consequent SSH-WFS measurements. Ideally the residual should be zero, however, detector and photon noise provide a noise floor in a realistic situation. Different color lines show the relation between full-well capacity utilization and the wavefront residual caused by noise.}
\label{f35s}
\end{figure}

Before measuring the relative wavefront, a number of frames are typically stacked to increase the effective exposure time and increase the precision of the final measurements. Figure~\ref{f35s} shows the RMS wavefront residual for a different number of stacked exposures and different source intensities. The results of this test demonstrated that the precision of the device is better than the required 5~nm across the pupil for any number of stacked images higher than 15 for the source intensity corresponding to the full well capacity of the detector. This is equivalent to 0.6 second of data acquisition for each data set which is also less than 5 second defined by our requirement on data acquisition time.

\section{Summary and Conclusion}
In this study, we introduced the optomechanical design of a low-cost, high-speed, very high-order Shack-Hartmann sensor and evaluated its capabilities, performance and noise characteristics. 

This detailed presentation of the optomechanical design provides insights into the mechanics and optics that govern the operation of the SSH-WFS. This helps in understanding the operational nuances of the system but also forms a blueprint for potential future enhancements or replications.

The performance metrics show that the SSH-WFS system is comparable to the HASO4 Shack-Hartmann device in terms of stability and linearity of measurements in lab conditions, while providing significantly more resolution. 

Our exploration into the noise characteristics of the system highlight the significance of the number of stacked exposures and the signal-to-noise ratio in determining measurement precision. The findings underscored that enhancing the effective exposure time, by stacking frames, can substantially augment the precision of the measurements. Moreover, the SSH-WFS's ability to maintain less than 5 nm precision across the pupil for stacked image counts exceeding 15, stands as a testament to its robustness.


\printbibliography 
\end{document}